\documentclass[preprint,12pt]{elsarticle}

\usepackage{amssymb}
\usepackage{amsmath}
\usepackage{xcolor}
\usepackage{float}
\journal{Optical Materials}

\begin{document}

\begin{frontmatter}


\title{Infrared Dielectric Function of Photochromic Thiazolothiazole Embedded Polymer\tnoteref{label1}}
%\tnotetext[label1]{}
\author[physics]{Nuren Z. Shuchi}
\ead{nshuchi@charlotte.edu}
\author[chemistry]{Tyler J. Adams}
\author[chemistry]{Naz F. Tumpa} %% Author name
\author[physics]{Dustin Louisos}
\author[physics]{Glenn D. Boreman}
\author[chemistry]{Michael G. Walter}
\author[physics-NJIT]{Tino Hofmann}

\affiliation[physics]{organization={Department of Physics and Optical Science, University of North Carolina at Charlotte}, addressline={9201 University City Blvd},
             city={Charlotte},
             postcode={28223},
             state={NC},
             country={U.S.A.}}
\affiliation[chemistry]{organization={Department of Chemistry, University of North Carolina at Charlotte}, addressline={9201 University City Blvd},
             city={Charlotte},
             postcode={28223},
             state={NC},
             country={U.S.A.}}             
\affiliation[physics-NJIT]{organization={Department of Physics, New Jersey Institute of Technology},
addressline={University Heights},
            city={Newark},
            postcode={07102},
            state={NJ},
            country={U.S.A.}}

\title{}



%% Abstract
\begin{abstract}
In this paper, the infrared dielectric function of photochromic dipyridinium thiazolo[5,4-d]thiazole embedded in polymer is reported. Bulk thiazolo[5,4-d]thiazole-embedded polymer samples were prepared by drop casting and dehydration in room temperature. The samples were investigated using spectroscopic ellipsometry before and after irradiation with a 405~nm diode laser in the infrared spectral range from 500~cm$^{-1}$ to 1800~cm$^{-1}$. The model dielectric functions of the thiazolothiazole embedded polymer film for its TTz$^{2+}$ (unirradiated) and TTz$^{0}$ (irradiated) states are composed of a series of Lorentz oscillators in the measured spectral range. A comparison of the obtained complex dielectric functions for the TTz$^{2+}$ and TTz$^{0}$ states shows that the oscillators located in the spectral ranges 500~cm$^{-1}$- 700~cm$^{-1}$, 1300~cm$^{-1}$- 1400~cm$^{-1}$, and 1500~cm$^{-1}$- 1700~cm$^{-1}$ change in both amplitude and resonant frequency upon transition between the states. Additionally, a resonance at approximately 1050~cm$^{-1}$ exhibited a change in oscillator amplitude but not resonant frequency due to the photochromic transition. 
\end{abstract}


% \begin{highlights}
% \item First report on the complex infrared model optical dielectric function of photochromic thiazolo[5,4-d]thiazole-embedded polymer from 500~cm$^{-1}$ to 1800~cm$^{-1}$ for the unirradiated and irradiated states
% \item Identified resonances in the spectral ranges 500~cm$^{-1}$- 700~cm$^{-1}$, 1300~cm$^{-1}$- 1400~cm$^{-1}$, and 1500~cm$^{-1}$- 1700~cm$^{-1}$ that change in both amplitude and resonant frequency upon transition between the unirradiated and irradiated states
% \item A resonance at approximately 1050~cm$^{-1}$ only changes in oscillator amplitude due to the photochromic transition but the resonant frequency remains constant 
% \end{highlights}

% %% Keywords
% \begin{keyword}
% %% keywords here, in the form: keyword \sep keyword

% %% PACS codes here, in the form: \PACS code \sep code

% %% MSC codes here, in the form: \MSC code \sep code
% %% or \MSC[2008] code \sep code (2000 is the default)

% \end{keyword}

\end{frontmatter}



%%%%%%%%%%%%%%%%%%%%%%%%%%  body  %%%%%%%%%%%%%%%%%%%%%%%%%%
\section{Introduction}
 Photochromism refers to the reversible transformation of a material between two states with distinct light absorption properties in different spectral regions induced by electromagnetic radiation \cite{crano1999organic}. 
Organic and inorganic photochromic materials have gained significant attention in recent years because of their potential applications in diverse fields. These applications include tinted lenses and smart windows, memory devices, actuators, tunable filters, and holographic gratings \cite{Tinted_eyewear,Cho_kim_MemoryDevice,Actuator,Tunable_filter}. Organic photochromic materials are emerging as more promising candidates compared to their inorganic counterparts such as metal halides \cite{MetalHalide_XinLi_2021} and transition metal oxides \cite{TMO_meng2024} 
for applications demanding spectral tunability, lower processing cost, and mechanical flexibility \cite{clark2010organic}.

Viologens constitute a significant class of organic photochromic materials \cite{guo2024_Viologen}. The photochromic behavior of viologens arises from photoinduced electron transfer. This transfer occurs in the polyvinyl alcohol/borax (PVA/borax) polymer matrix to the pyridinium cation, resulting in the reduction of the pyridinium unit and the formation of a radical cation \cite{Viologen_mechanism}. 

Viologens can be extended with a thiazolo[5,4-d]thiazole (TTz) fused, conjugated bridge. This approach has been attracting increasing interest due to their the strong fluorescence, solution-processability, and reversible photochromic transition. In particular, dipyridinium thiazolo[5,4-d]thiazole viologens exhibit high-contrast, fast, and reversible photochromic changes when embedded in a polymer matrix. When exposed to radiation with an energy larger than 2.8~eV, it transitions from light yellow (TTz$^{2+}$) to purple (TTz$^{\cdot+}$) to blue (TTz$^{0}$) state due to two distinct photoinduced single electron reductions. The reverse color change (to yellow/colorless) is driven by reaction of the TTz$^{0}$ state with molecular oxygen \cite{adams2021}. 

The accurate knowledge of the dielectric function is essential to contemporary research on the design, fabrication and optimization of tunable optical devices utilizing photochromic thiazolothiazole-embedded polymers. In particular finite element numerical calculations rely on dielectric function data for the simulation of complex structures and devices. % Further advances in the design and fabrication of TTz-based optically tunable devices are currently impaired by a gap in accurate knowledge of the complex dielectric function of the material. 
We have previously addressed this knowledge gap and reported on the complex dielectric function of a nonphotochromic TTz derivative and a photochromic TTz-embedded polymer in the visible and near-infrared spectral range \cite{NShuchi2023_TTz,Tadams2024_TTz_2024}. The infrared spectral range, however, has not yet been investigated. 

In this paper, we report on a parameterized dielectric function of photochromic dipyridinium thiazolo[5,4-d]thiazole embedded in polymer obtained from a quantitative analysis of the polarization-sensitive optical response. The measurements were taken in the infrared spectral range from 500~cm$^{-1}$ to 1800~cm$^{-1}$ before and after irradiation with a 405~nm diode laser as an excitation source. The model dielectric functions of the thiazolothiazole embedded polymer film for its TTz$^{2+}$ (unirradiated) and TTz$^{0}$ (irradiated) states are composed of a series of Lorentz oscillators in the measured spectral range. A comparison of the obtained complex dielectric functions for the TTz$^{2+}$ and TTz$^{0}$ state shows several infrared absorption bands for which both amplitude and resonant frequency change upon transition between the states. In addition, a resonance has been identified at approximately 1050~cm$^{-1}$, for which, only a change of the oscillator amplitude was observed due to the photochromic transition.

In addition to providing the infrared dielectric function for thiazolothiazole-embedded polymer in the oxidized and reduced state, the parameterized dielectric function of PVA/borax is reported. This provides a comparison between the thiazolothiazole-embedded polymer and the PVA/borax host-polymer.

% In this paper, we report on the infrared dielectric function of photochromic dipyridinium thiazolo[5,4-d]thiazole embedded in polymer determined by spectroscopic ellipsometry. The measurements were taken in the infrared spectral range from 500~cm$^{-1}$ to 1800~cm$^{-1}$ before and after irradiation with a 405~nm diode laser as an excitation source. The model dielectric functions of the thiazolothiazole embedded polymer film for its TTz$^{2+}$ (unirradiated) and TTz$^{0}$ (irradiated) states are composed of a series of Lorentz oscillators in the measured spectral range. A comparison of the obtained complex dielectric functions for the TTz$^{2+}$ and TTz$^{0}$ state shows several infrared absorption bands for which both amplitude and resonant frequency change upon transition between the states. In addition, a resonance has been identified at approximately 1050~cm$^{-1}$, for which, only a change of the oscillator amplitude was observed due to the photochromic transition. 

\section{Experiment}

\subsection{Synthesis and sample preparation}
Dipyridinium thiazolo[5,4-d]thiazole was synthesized by refluxing dithiooxamide and 4-pyridinecar\-boxaldehyde in dimethylformamide (DMF) at 153$^{\circ}$C for 8 hours. The resulting dipyridal TTz was then heated with 3-bromopropyl trimethylammonium bromide in DMF to alkylate the pyridine rings and increase its water solubility. Further details on the  synthesis of dipyridinium TTz can be found in our previous publications\cite{adams2021,Tadams2024_TTz_2024}. Following the synthesis 3.4 wt\% dipyridinium TTz was dissolved in polyvinyl alcohol solution and then borax is added into the mixture to obtain a TTz embedded polymer hydrogel as previously reported \cite{adams2021}.  

Bulk samples with a thickness of approximately 150~\textmu m were prepared by drop casting and dehydration at room temperature. In addition to the TTz embedded polymer sample, a reference sample composed of a polymeric system based on polyvinyl alcohol and borax (PVA/borax) without dipyridinium TTz was prepared using the same approach.

\subsection{Data acquisition and analysis}

The reference and bulk TTz-embedded polymer samples were investigated using a commercial infrared ellipsometer (Mark I IR-VASE, J.A. Woollam Company). This ellipsometer operates in a rotating polarizer - sample - rotating compensator - rotating analyzer configuration as described in Ref.~\cite{Fujiwara2007} and employs a Boman FTIR spectrometer and a DTGS detector. A 405~nm diode laser was used as an excitation source to investigate the optical properties of the TTz$^{0}$ state of the TTz-embedded polymer. The ellipsometric $\varPsi$- and $\varDelta$-spectra were obtained before and after irradiation in the infrared spectral range from 500~cm$^{-1}$ to 1800~cm$^{-1}$ with a resolution of 8~cm$^{-1}$ for a single angle of incidence $\varPhi_a$=65$^{\circ}$. All measurements were carried out at room temperature in a nitrogen atmosphere. The nitrogen atmosphere prevents unintended transitions from the TTz$^{0}$ to the TTz$^{2+}$ state due to oxidation during data acquisition.

The stratified optical layer calculations required for the analysis of the spectroscopic ellipsometry data were carried out using a commercial software package (WVASE32, J.A. Woollam Co. Inc.). For these calculations parameterized Kramers-Kronig consistent dispersion functions with sufficient flexibility to accurately reproduce the optical features of the experimental $\varPsi$- and $\varDelta$-spectra (see Fig.~\ref{Blue_and_Yellow}) must be employed.

Sums of oscillators with Lorentzian broadening are widely used to describe the signatures of molecular or lattice vibrations in the infrared dielectric function \cite{Synowicki2004}.  This approach ensures that the extracted complex model dielectric function is Kramers-Kronig consistent. The use of parameterized dielectric function models offers an advantage over a point-by-point based analysis approach where $\varepsilon_1$ and $\varepsilon_2$ are determined independently for each wavelength. The former prevents experimental noise from entering the model dielectric function and ensures Kramers-Kronig consistency.

For the infrared model dielectric functions of both the reference sample and the TTz-embedded polymer in the TTz$^{2+}$ and TTz$^{0}$ states it is found that a sum of oscillators with Lorentzian broadening describes the optical response well: 
\begin{eqnarray} \label{eqn:1}
\varepsilon(E)&=&\varepsilon_1(E)+i\varepsilon_2(E)\ ,\\\nonumber
&=&\varepsilon_{\infty}+\sum_{k=1}\frac{A_k\gamma_kE_k}{E_k^2-E^2-i\gamma_k E}\  ,
\end{eqnarray}

\noindent where $\varepsilon_1(E)$ and $\varepsilon_2(E)$ are the real and imaginary parts of the complex dielectric function, respectively, as a function of the photon energy $E$. The relevant physical parameters $A_k$, $E_k$, and $\gamma_k$ in the oscillator functions represent the oscillator amplitude, resonant energy, and broadening, respectively. The reference sample's dielectric function was adequately modeled with twelve Lorentz oscillators. The parameterized model dielectric function of the TTz-embedded polymer required eleven Lorentz oscillators to describe TTz$^{2+}$ and TTz$^{0}$ states of the TTz-embedded polymer. Relevant parameters, such as oscillator amplitude, resonant energy, and broadening, were varied during the analysis using a Levenberg-Marquardt algorithm to minimize the weighted error function $\chi^2$, and achieve the best fit between experimental and calculated data \cite{JellisonJr2000TSF,JellisonJr1993_MSE}.

\section{Results and Discussion}

\begin{figure}[H]
%\centering

\includegraphics[width=\linewidth]{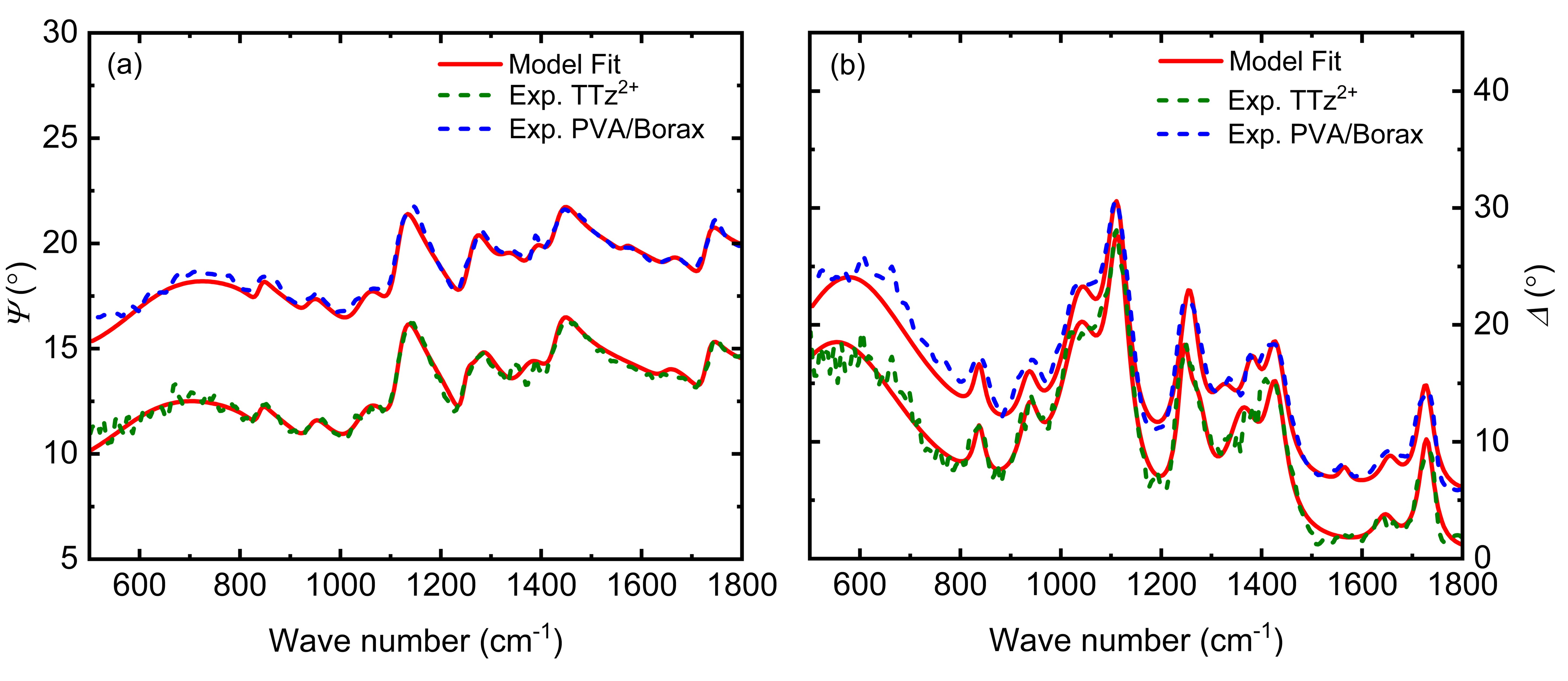}
\caption{\label{PVA-Yellow}Experimental (dashed lines) and best-model calculated (red solid lines) $\varPsi$-spectra (a) and $\varDelta$-spectra (b), respectively, for the TTz$^{2+}$ (green dashed lines) state of the bulk TTz-embedded polymer and the PVA/Borax reference sample (blue dashed lines). The  spectra were obtained at a single angle of incidence $\varPhi_{\rm a}= 65^{\circ}$ under nitrogen atmosphere at room temperature. The spectra for the PVA/Borax reference sample are depicted with an offset of $5^{\circ}$ for visibility.}
\end{figure}

Fig.~\ref{PVA-Yellow} illustrates the experimental (dashed lines)  and best-model calculated (red solid lines) $\varPsi$-spectra (a) and $\varDelta$-spectra (b), respectively, for the TTz-embedded polymer in its TTz$^{2+}$ (green dashed lines) in comparison to the PVA/borax reference sample (blue dashed lines). The experimental data were acquired at an angle of incidence $\varPhi_{\rm a}= 65^{\circ}$ in a nitrogen environment. For visibility, the experimental and best-model calculated $\varPsi$- and $\varDelta$-spectra for the PVA/borax reference sample are shown with an offset of $5^{\circ}$ in Fig.~\ref{PVA-Yellow}. The similarity between the spectra for the PVA/borax reference sample and the TTz-embedded polymer in its TTz$^{2+}$ state can be attributed to the moderate concentration of TTz (3.4 wt\%) within the PVA/borax matrix. However, subtle differences in the ellipsometric data can be observed in the spectral regions of 1300-1400~cm$^{-1}$ and 1500-1700~cm$^{-1}$. Especially, the absorption peaks observed in the PVA/borax reference at approximately 1327~cm$^{-1}$  and 1562~cm$^{-1}$ are suppressed in the spectra of the TTz-embedded polymer in its TTz$^{2+}$ state. This suppression in the aforementioned spectral regions is associated with the presence of 3.4 wt\% dipyridinium TTz in the polymer matrix.

\begin{figure}[H]
%\centering

\includegraphics[width=\linewidth]{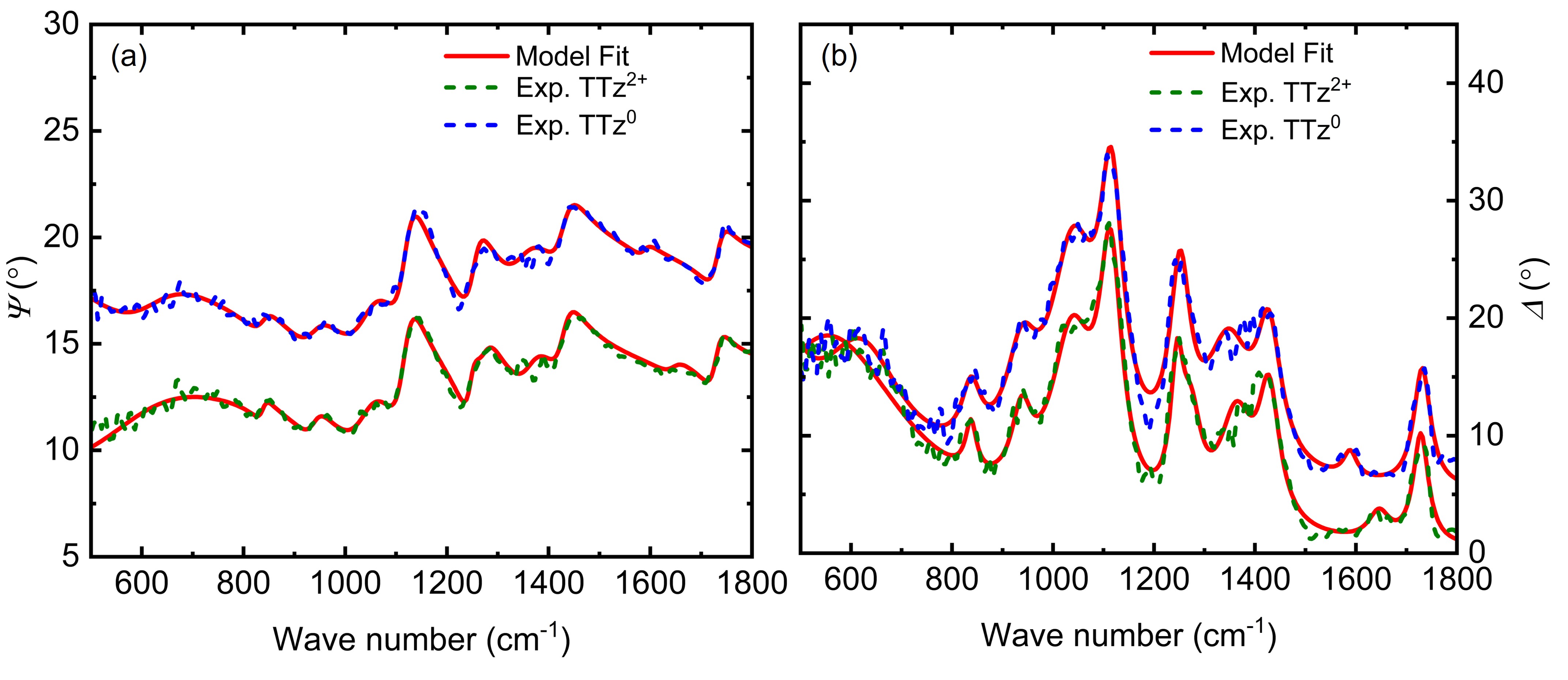}
\caption{\label{Blue_and_Yellow}  Experimental (dashed lines) and best-model calculated (red solid lines) $\varPsi$-spectra (a) and $\varDelta$-spectra (b), respectively, for TTz$^{2+}$ (green dashed lines) and TTz$^{0}$ (blue dashed lines) states of the bulk TTz-embedded polymer sample. These spectra were obtained at a single angle of incidence $\varPhi_{\rm a}= 65^{\circ}$ under nitrogen atmosphere at room temperature. The spectra for TTz$^{0}$ state are depicted with an offset of $5^{\circ}$ for visibility.}
\end{figure}

Fig.~\ref{Blue_and_Yellow} depicts the experimental (dashed lines)  and best-model calculated (red solid lines) $\varPsi$-spectra (a) and $\varDelta$-spectra (b), respectively, for the TTz-embedded polymer in its TTz$^{2+}$ (green dashed lines) and TTz$^{0}$ states (blue dashed lines). For visibility, the experimental and best-model calculated $\varPsi$- and $\varDelta$-spectra for the bulk TTz sample in the TTz$^{0}$ state are plotted with an offset of $5^{\circ}$ in Fig.~\ref{Blue_and_Yellow}.

The experimental and best-model calculated $\varPsi$- and $\varDelta$-spectra are in good agreement over the entire investigated spectral range in which a total of eleven vibration bands can be observed. The best-model fits yielded $\chi^2$ values of 0.6075, and 0.6703 for the TTz$^{2+}$ and TTz$^{0}$ states, respectively. These values indicate good agreement between the experimental and best-model calculated data over the measured spectral range \cite{JellisonJr1993_MSE}.

\begin{figure}[hbt]
%\centering
\includegraphics[width=\linewidth]{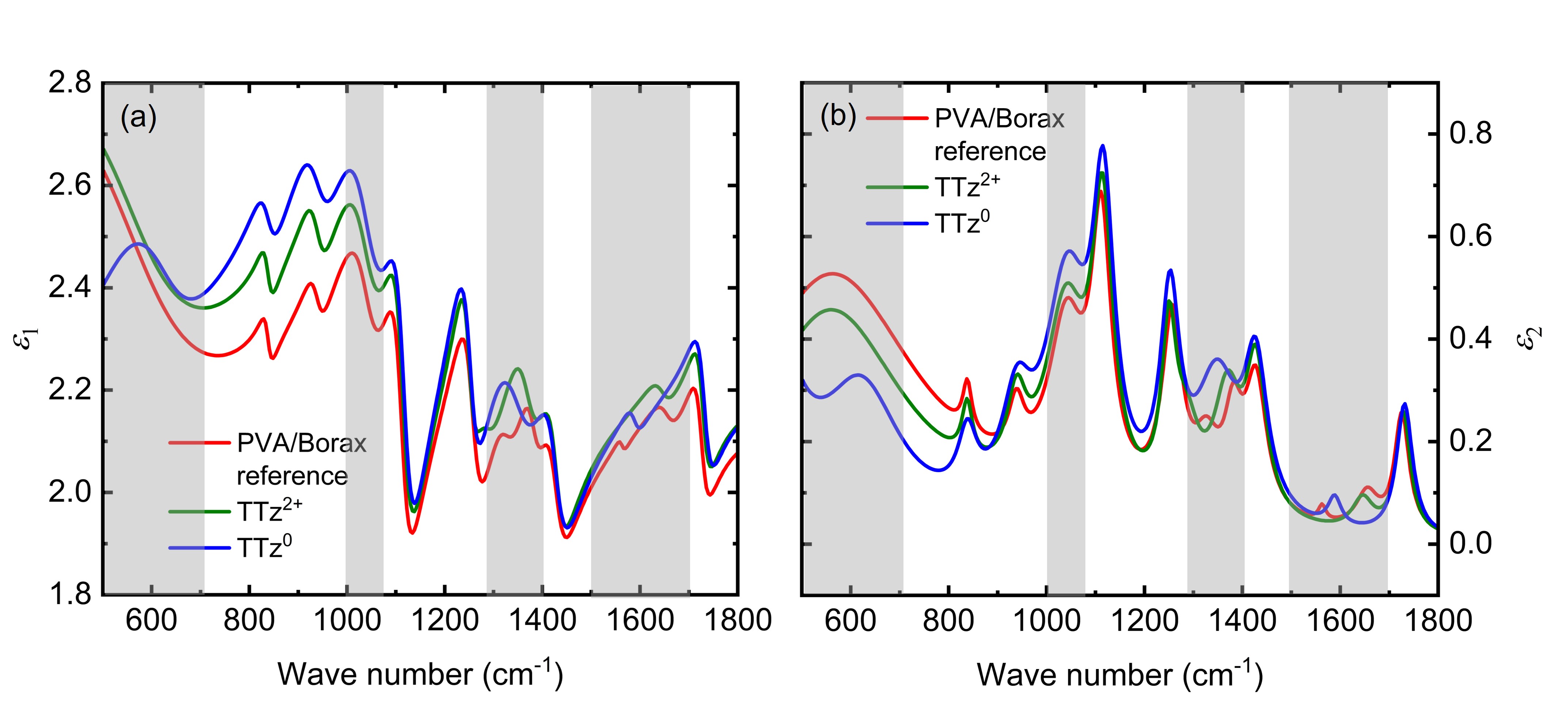}
\caption{\label{comparison} Comparison between the real $\varepsilon_1$ (a) and imaginary $\varepsilon_2$ parts (b) of the best-fit parameterized model dielectric functions for PVA/Borax (red solid lines), TTz$^{2+}$ (green solid lines), and TTz$^{0}$ (blue solid lines) states. The shaded regions illustrate the spectral windows where photochromically induced changes in the infrared dielectric function of the TTz-embedded polymer occur.}
\end{figure}

Due to the moderate concentration of TTz within the PVA/borax matrix, the majority of the observed infrared absorption bands can be attributed to PVA/borax (see also Fig.~\ref{comparison}). 
%TO-DO is this statement correct?
It is known that the vibration bands observed at approximately 847~cm$^{-1}$ and 1104~cm$^{-1}$ can be attributed to the symmetric and asymmetric stretching vibrations of B$_4$\textendash O, respectively \cite{PVA_borax_oliveira2009,PVA_borax_dave2018,lawrence2024_PVA}. Additional resonances are observed  approximately at 1417~cm$^{-1}$ and 1661~cm$^{-1}$. These resonances correspond to asymmetric stretching relaxation of B\textendash O\textendash C and the bending vibration of HOH in water molecules, respectively \cite{lawrence2024_PVA}. 

In contrast to vibrational bands attributed to PVA/borax, the photo\-chro\-mically-induced, infrared-optical changes to the TTz viologen can be observed only as very subtle features. The most prominent photochromic changes can be noticed at approximately 1600~cm$^{-1}$ and below 750~cm$^{-1}$. Additional photochromic changes occur at approximately 1080~cm$^{-1}$ and between 1300 and 1400~cm$^{-1}$.

As discussed above, the majority of the observed infrared absorption bands can be attributed to PVA/borax. This is corroborated by a direct comparison of the ellipsometric data of the PVA/borax reference sample to the TTz-embedded polymer sample as shown in Fig.~\ref{PVA-Yellow}. 

%1 MDF
Figures \ref{comparison}(a) and \ref{comparison}(b) show the real $\varepsilon_1$ and imaginary $\varepsilon_2$ components of the best-fit parameterized model dielectric functions for the reference sample, and the TTz-embedded polymer sample in the  TTz$^{2+}$, and TTz$^{0}$ states. Shaded regions in these figures highlight the spectral regions where the dielectric function of the TTz-embedded polymer before and after irradiation with a 405~nm diode laser and dielectric function of the reference sample deviate from each other.

In the following section, we will focus primarily on the imaginary part of the dielectric function to discuss the observed infrared absorption bands. As can be seen in Fig.~\ref{comparison}(b) the main infrared absorption bands are due to PVA/Borax (red solid lines). These bands include the symmetric and asymmetric stretching vibrations of B$_4$\textendash O, asymmetric stretching relaxation of B\textendash O\textendash C and the bending vibration of HOH in water molecules \cite{PVA_borax_oliveira2009,PVA_borax_dave2018,lawrence2024_PVA}.

The imaginary part $\varepsilon_2$ of infrared dielectric function of the reference sample, within the measured spectral range, exhibits several prominent absorption peaks as shown in Fig.~\ref{comparison}(b). The dielectric response for the TTz$^{2+}$ state of the TTz-embedded polymer closely resembles the absorption features of the reference sample. This similarity in optical response is a consequence of the low concentration of dipyridinium TTz incorporated into the cross-linked PVA/borax polymer. However, distinct spectral differences are also evident in Fig.~\ref{comparison}(b). Notably, in the spectra of the TTz-embedded polymer in its TTz$^{2+}$ state, the absorption peaks seen in the PVA/borax reference at approximately 1327~cm$^{-1}$  and 1562~cm$^{-1}$ are suppressed. This suppression in the 1300–1400~cm$^{-1}$ and 1500–1700~cm$^{-1}$ spectral regions is attributed to the presence of 3.4 wt\% dipyridinium TTz in the polymer matrix. Furthermore, a comparison of the complex dielectric functions obtained for the TTz-embedded polymer before and after irradiation reveals that oscillators in the spectral ranges of 500–700 cm$^{-1}$, 1300–1400 cm$^{-1}$, and 1500–1700 cm$^{-1}$ exhibit changes in both amplitude and resonant frequency upon transition between these states. Additionally, a resonance near 1050 cm$^{-1}$ was identified, where only an amplitude change was observed due to the photochromic transition.

\section{Conclusion}
In conclusion, the first quantitative analysis of the complex infrared optical dielectric function of photochromic thiazolo[5,4-d]thiazole-embedded polymer using spectroscopic ellipsometry is reported. Spectroscopic ellipsometry is the preeminent technology for the accurate determination of dielectric material responses. However, while spectroscopic ellipsometry has been used in the visible spectral range to investigate a wide range of photochromic materials, the direct ellipsometric investigation of photochromic materials in the infrared spectral range has been lacking \cite{PSnyder}.

In contrast to infrared reflection or transmission measurements which are frequently employed to investigate the infrared optical properties of photochromic polymers, our approach ensures the Kramers-Kronig consistency of the dielectric function. The accuracy of the dielectric function is an essential foundation for further numerical first-principle analysis of the material in order to elucidate the nature of the photochromic transitions. Currently the nature of these transitions is not completely understood, which hinders the further development of thiazolothiazole-based photochromic materials. 
 
The fabricated bulk TTz-embedded polymer samples were studied before and after irradiation with a 405~nm diode laser in the spectral range from 500~cm$^{-1}$ to 1800~cm$^{-1}$. To describe the optical features of the TTz-embedded polymer accurately, complex-valued model dielectric functions composed of Lorentz oscillators were used for both states. A comparative study of the obtained complex dielectric functions for the TTz$^{2+}$ and TTz$^{0}$ states reveals that the oscillators located in the spectral ranges 500~cm$^{-1}$- 700~cm$^{-1}$, 1300~cm$^{-1}$- 1400~cm$^{-1}$, and 1500~cm$^{-1}$- 1700~cm$^{-1}$ change in both amplitude and resonant frequency upon transition between the states. 

Furthermore, the a resonance near 1050~cm$^{-1}$ was identified that exhibited a change in amplitude, but not in resonant frequency, following the photochromic transition. By providing a detailed analysis of the TTz polymer's optical transitions in the spectral range from 500 cm$^{-1}$ to 1800 cm$^{-1}$  before and after irradiation, the work reported here establishes a foundation for integrating TTz-based materials into infrared-tunable optical devices.

The findings advance the field by demonstrating how photo-induced structural changes influence the dielectric response in the infrared, which is critical for the development of tunable infrared optics and metamaterials. By bridging this gap, our work paves the way for future innovations in dynamic and adaptive infrared optical systems.

\section*{Funding}
The authors acknowledge the support from the National Science Foundation within the IUCRC Center for Metamaterials (2052745) and National Science Foundation Grant (CHE-2400165).

\section*{Acknowledgment}
The authors would like to acknowledge the support from the Department of Physics and Optical Science and the Department of Chemistry at the University of North Carolina at Charlotte. We further acknowledge support from the Center for Optoelectronics and Optical Communications, the Division of Research, and the Klein College of Science at UNC Charlotte.

%\appendix
%\section{Example Appendix Section}
%\label{app1}


%%       \cite{<label>} ==> [1]

%%
%Example citation, See \cite{lamport94}.




%%%%%%%%%% If using BibTeX:
%\bibliographystyle{elsarticle-num}
%\bibliography{Photochromic_TTz_IR}

\end{document}